\documentclass{PoS}

\usepackage{graphicx}
\usepackage{color}
\usepackage{epsfig,mathbbol}
\newcommand{\preprintline}{\newline
\vskip -4.2cm
\rightline{\parbox{4cm}{\large\rm HU-EP-05/53 \\ ITEP-LAT 2005-21}}
\vspace{3.2cm}}

\PoS{PoS(LAT2005)306}

\title{Topological clusters in SU(2) gluodynamics at finite temperature and 
the evidence for KvB calorons\preprintline}

\ShortTitle{Topological clusters and KvB calorons}

\author{\speaker{Ernst-Michael Ilgenfritz}\\
Institut f\"ur Physik, Humboldt Universit\"at zu Berlin,
12489 Berlin, Germany\\
E-mail: \email{ilgenfri@physik.hu-berlin.de}}

\author{Philipp Gerhold and Michael M\"uller-Preussker\\
Institut f\"ur Physik, Humboldt Universit\"at zu Berlin,
12489 Berlin, Germany\\
E-mail: \email{gerhold@physik.hu-berlin.de, mmp@physik.hu-berlin.de}}

\author{B.~V.~Martemyanov and A.~I.~Veselov\\
ITEP, Bolshaya Cheremushkinskaya 25, 117259 Moscow, Russia \\
E-mail: \email{martemja@itep.ru, veselov@itep.ru} }

\abstract{
We report on our search for Kraan-van Baal calorons in finite temperature
SU(2) lattice ensembles. We also discuss recent progress made in developing 
a caloron-anticaloron gas model decribing confinement and deconfinement in 
the context of trivial and non-trivial holonomy.  }

\FullConference{XXIIIrd International Symposium on Lattice Field Theory\\
                 25-30 July 2005\\
                 Trinity College, Dublin, Ireland}

\begin{document}

\vspace{-2mm}
\section{Introduction}
With the help of overlap fermions, we understand better today 
how the topological charge is distributed in the vacuum,
seen without or with an UV filter~\cite{horvath,koma,weinberg}. 
Still, there is a quest for a picture in terms of semiclassical 
gluonic field excitations which can be held responsible 
for the non-perturbative properties of QCD.
Indeed, a locally (anti)selfdual background becomes visible under 
moderate smearing of a lattice gauge fields taken from the confinement 
phase~\cite{degrand}. A structure like this is also what is seen 
by low-lying fermion modes~\cite{koma,weinberg}. 
The instanton liquid model~\cite{shuryak} (built with calorons, at $T \ne 0$) 
seems to describe the background fields as superpositions
of semiclassical objects. 
Confinement survives controlled smearing~\cite{degrand}. Therefore 
it is a challenge to describe the string tension in a gas or liquid model.
With a liquid of instantons one fails to generate a linear potential 
between fundamental charges at $R >> \rho$ (instanton radius), and the 
static forces of different representations cannot be simultaneosly described, 
even at shorter distances.

After caloron solutions with non-trivial asymptotic holonomy have been
discovered~\cite{KvB} the hope was revived to connect topology and confinement 
at the semiclassical level. Two things are required. 
First one has to find, below some temperature $T_{\rm dec}$, that the effective 
potential of the Polyakov loop, generated by a gas of (anti)calorons,
develops a minimum at zero~\cite{diakonov}. Second, one should be able to 
demonstrate that the caloron gas creates a non-vanishing (temporal) string 
tension, {\it e.g.} from the Polyakov loop correlator at finite temperature, 
if the asymptotic Polyakov loop is vanishing (confinement). The spatial string 
tension, derivable from spatial Wilson loops, should remain unaltered if the
asymptotic Polyakov loop gets non-vanishing (deconfinement), and it should be 
rising at high temperature. 

We will present preliminary results supporting such a scenario in the second 
part of this report. Before that, we will update our recent search 
for KvB caloron-like features~\cite{clusterKvB} in Monte Carlo configurations 
analyzed after smearing. This type of analysis has been extended now to 
various temperatures below $T_{dec}$.

\vspace{-2mm}
\section{SU(2) Calorons at high and low temperature}

\begin{figure}[!htb]
\centering
\includegraphics[width=0.35\textwidth]{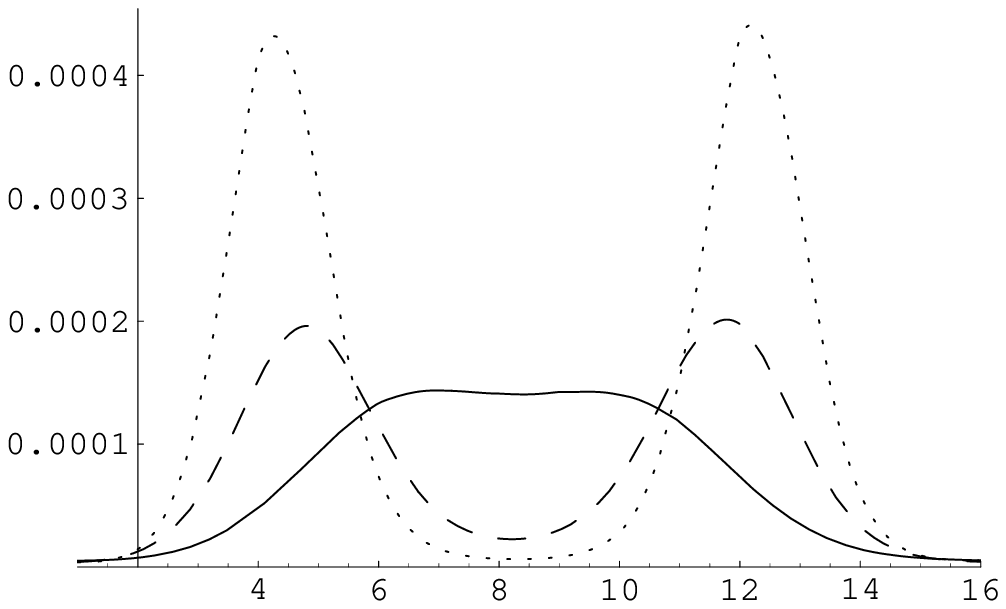}%
\hspace{0.5 cm}
\includegraphics[width=0.35\textwidth]{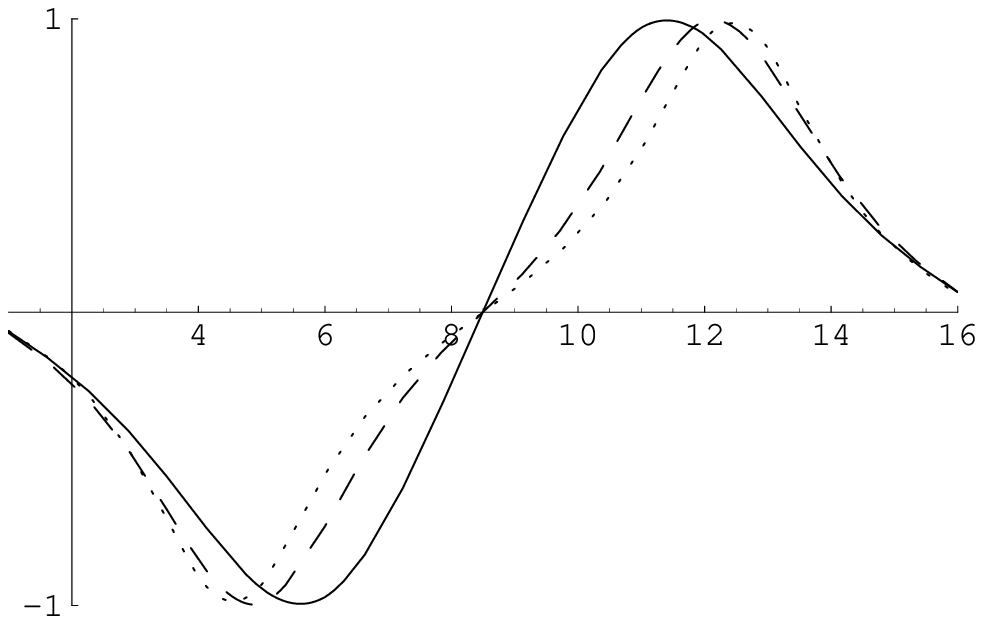}
\caption{Action density (left) and Polyakov loop (right) along the axis 
between constituents at fixed distance in a SU(2) caloron for different
temperatures. At lower $T$ the lumps coalesce whereas the peaks of the 
Polyakov loop still localize the constituents.~\cite{epsilon} }
\label{fig:fig1}
\end{figure}

The distinguishing feature of the KvB caloron~\cite{KvB} is the existence of 
$N_{\mathrm color}$ monopole constituents. Let $d$ be the distance between
the two, for SU(2).
The size of the constituent lumps is ${\cal O}(1/\pi T)$. 
They share the action and topological charge in a proportion 
$\omega/(\frac{1}{2}-\omega)$ determined by the {\it asymptotic holonomy}, 
{\it i.e.} the Polyakov loop $L_{\infty}=\cos (2 \pi \omega)$). 
If the constituents are well-separated relative to their size they 
form static lumps of action and charge. They 
are localizable by degenerate eigenvalues of the {\it local holonomy},
meaning $L({\vec x}) \to \pm 1$ for SU(2) (see Fig. \ref{fig:fig1}).
In the maximal Abelian gauge (MAG), Abelian monopoles are emerging from
Abelian projection that can serve to localize the 
(otherwise non-Abelian and gauge independent !) monopoles. 
Being sources of magnetic and electric fields (${\vec B}^a = \pm {\vec E}^a$), 
they are called dyons. 
At low temperature, a parameter $\rho$ (with $\rho^2 = d/(\pi T)$)
plays the role of the size of a single (instanton-like) lump.

The elusive fractional topological charge constituents~\cite{epsilon} have 
also been seen using overimproved cooling on the symmetric torus, 
appearing as parts of a classical 
solution, with a size ${\cal O}(L)$ determined solely by the volume. 
In Monte Carlo (equilibrium) configurations such lumps get more localized 
as the result of quantum fluctuations. Such instanton constituents might 
have been detected on the symmetric torus 
already with the help of fermionic zero modes~\cite{gattringer} 
which jump under the change of boundary conditions (periodic/antiperiodic).
This interpretation is under scrutiny now together with these authors.
\vspace{-2mm}

\section{Caloron-like clusters ?~\cite{clusterKvB}}

We have examined the Abelian monopole content of topological clusters 
which become visible in smeared configurations. The aim was to identify 
lumps of charge, which are likely to be ``undissociated'' calorons, and
those which are isolated dyons (single monopoles accompanied 
by topological charge). This study has been performed on 500 
configurations on a $20^3\times6$ lattice created at (Wilson) $\beta=2.3$.  
The configurations were subjected to $50 \ldots 100$ steps of 4D APE-smearing. 
A cluster analysis was then searching for connected clusters of lattice 
points with charge density $|q(x)| > q_{threshold}$. 
We have adapted the threshold to give a maximal number of separate 
clusters.

The distribution of the local Polyakov loop at sites passed by timelike 
Abelian monopoles was found very different from the unbiased local distribution. 
Then we considered topological clusters according to the monopoles detected inside.
We should mention that many topological clusters remain unclassified in this way. 
The monopole lines occur either as isolated static monopoles, 
as monopole-antimonopole ($M\bar{M}$) pairs 
or as closed monopole trajectories. On this basis we have
classified the clusters containing monopoles. We were mainly interested in two
cluster observables. One is the averaged Polyakov loop 
$\langle PL({\rm Abelian~monopoles}) \rangle_{\rm cluster}$ 
(averaged over sites near timelike monopoles inside the clusters). 
In clusters with $M\bar{M}$ or with closed monopole trajectories this average 
turns out close to zero, for isolated monopoles the distribution of this
observable is peaked far from zero (around $\pm 0.75$).
The other cluster observable is the cluster topological charge 
$Q_{\rm cluster}$, 
assigned to the cluster by summing up the topological density within a
radius estimated from the maximal density in the cluster. 
This estimate is model dependent, taking the type of cluster into account.
For details we refer to Ref.~\cite{clusterKvB}. 
\begin{figure}[h]
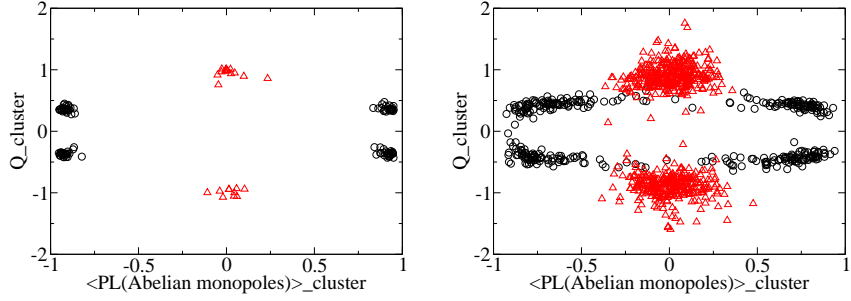

\centering
\includegraphics[width=0.35\textwidth]{CLUSTER.fig6b.eps}%
\hspace{0.5 cm}
\includegraphics[width=0.35\textwidth]{CLUSTER.fig4b.eps}
\caption{Scatterplots of topological clusters forming calorons (left) 
and in smeared Monte Carlo configurations (right) showing dyons (black 
circles) and non-dissociated calorons (red triangles).}
\label{fig:fig2}
\end{figure}

As a test, the method has been applied to single calorons randomly created 
with some distribution of distances $d$ between the constituents. 
The resulting scatterplot is shown in Fig. \ref{fig:fig2} (left). 
The clustering towards 
$(Q_{\rm cluster},\langle PL({\rm Abelian~monopoles}) \rangle_{\rm cluster}) 
= (\pm 1, 0)$ and $(\pm \frac{1}{2}, \pm 1)$ reaffirms the method and the
usefulness of the cluster variables. 
The same scatterplot for smeared equilibrium configurations 
is shown in Fig. \ref{fig:fig2} (right).  
Both figures refer to MAG as the method to find the monopoles which are 
crucial for this analysis. We have also used the Polyakov gauge.
\begin{figure}[!htb]
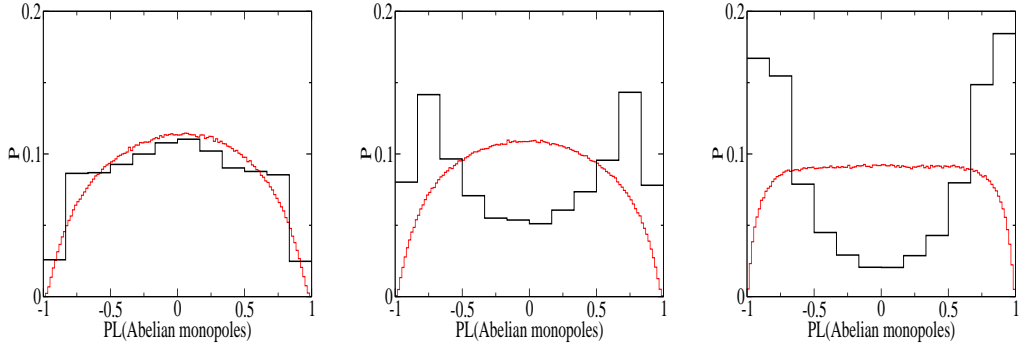

\centering
\includegraphics[width=.27\textwidth,height=.3\textwidth]{plmoncorr22.eps}%
\hspace{0.5 cm}
\includegraphics[width=.27\textwidth,height=.3\textwidth]{plmoncorr23.eps}%
\hspace{0.5 cm}
\includegraphics[width=.27\textwidth,height=.3\textwidth]{plmoncorr24.eps}
\caption{Distribution of Polyakov loops in sites with time-like Abelian 
monopoles (black thick line). For comparison the unbiased distribution of 
Polyakov loops in all sites is shown (red thin line).
The subpanels show this for $\beta = 2.2$ ,$2.3$, $2.4$ (from left 
to right).  }
\label{fig:fig3}
\end{figure}

Recently we have repeated this analysis with 200 configurations 
at (Wilson) $\beta=2.2$, $2.3$ and $2.4$ (corresponding to 
$T/T_{\rm dec}= 0.5$, $0.63$, and $0.88$) in the confined phase.
The lattice size was now $24^3\times6$, and 50 steps of 4D APE-smearing have been
applied. The MAG fixing has been refined (using now simulated annealing from
5 random start copies). Now we are in the position to see how the 
above picture depends on the temperature. 
As Fig. \ref{fig:fig3} shows, 
the correlation between Abelian monopoles and the Polyakov loop 
becomes stronger (weaker) with increasing (decreasing) $\beta$. 
\begin{figure}[!htb]
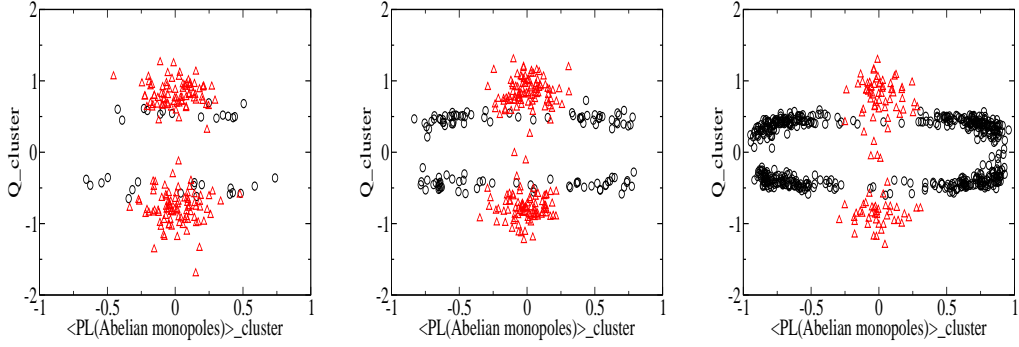

\centering
\includegraphics[width=.27\textwidth,height=.3\textwidth]{plqcorrclust22.eps}%
\hspace{0.5 cm}
\includegraphics[width=.27\textwidth,height=.3\textwidth]{plqcorrclust23new.eps}%
\hspace{0.5 cm}
\includegraphics[width=.27\textwidth,height=.3\textwidth]{plqcorrclust24.eps}\\
\caption{The Figure shows similar scatterplots as Fig. 2 b, now for 
$\beta = 2.2$, $2.3$, $2.4$ (from left to right) in the refined analysis (see text).} 
\label{fig:fig4}
\end{figure}

We have checked the contingency of the cluster classification according 
to monopoles in MAG and Polyakov gauge. 
Fig. \ref{fig:fig4} refers now only to clusters which are identically 
recognized in both gauges.
The clustering in these scatterplots should be considered as a hint
for the existence of calorons and dyons. It is obvious that the probability
to detect the dyon gas 
increases towards $T_{\rm dec}$. At lower temperature we expect no 
evidence for separate dyons. At low $T$ topological clusters tagged by 
their monopole content are mostly non-static lumps with a charge
clustering around $\pm 1$.  
We are still unable, applying this method to the deconfined phase, 
to understand the topological objects there in terms of KvB calorons.

\section{A SU(2) KvB caloron gas model}

We have constructed a model based on random superpositions of KvB calorons.
There are technical details that can be described only in a forthcoming 
publication: fixed holonomy boundary conditions, 
how to avoid interaction of calorons to be added with Dirac strings of 
existing calorons, improving overlapping calorons.
We consider an equal number of calorons and anticalorons randomly distributed 
in a continuous box. On an embedded open $32^3\times8$ lattice  
the gauge field is discretized for further analysis. 
The temperature (and lattice scale) was 
chosen setting $T=T_{\rm dec}$, the deconfinement temperature. 
With a density $n = 1 {\rm fm}^{-4}$, on average 64 (anti)calorons will occupy 
the lattice 4-volume. The aim is to demonstrate, in this conservative setting, 
the crucial role of the holonomy parameter $\omega$. 
Temperature $T$ and holonomy $\omega$ entering the KvB caloron solutions are
prescribed whereas the constituent positions $x_1$ and $x_2$ are random
(one of the time coordinates being equivalent to a global rotation). 
The 3D distance $d=|{\vec x}_1-{\vec x}_2|$ has been sampled according 
to a distribution obtained from  
$n(\rho) \propto \rho^{b-5} \exp (- c~\rho^2) \mbox{ , } b=11~N_{\rm color}/3$ 
to give an average $\bar{\rho} = 0.4$ fm.
This choice was suggested in order to remain close to the instanton model.
It turns out that in this case 
only a small fraction of calorons is really dissociated 
into dyons. 

\begin{figure}[!htb]
\centering
\includegraphics[width=7.0cm]{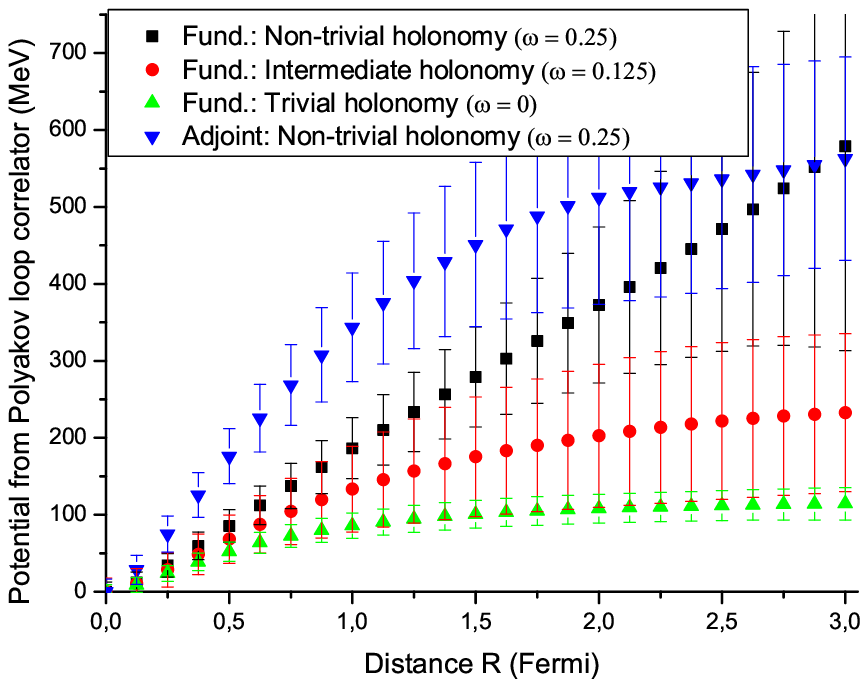}%
\hspace{0.1 cm}
\includegraphics[width=7.0cm]{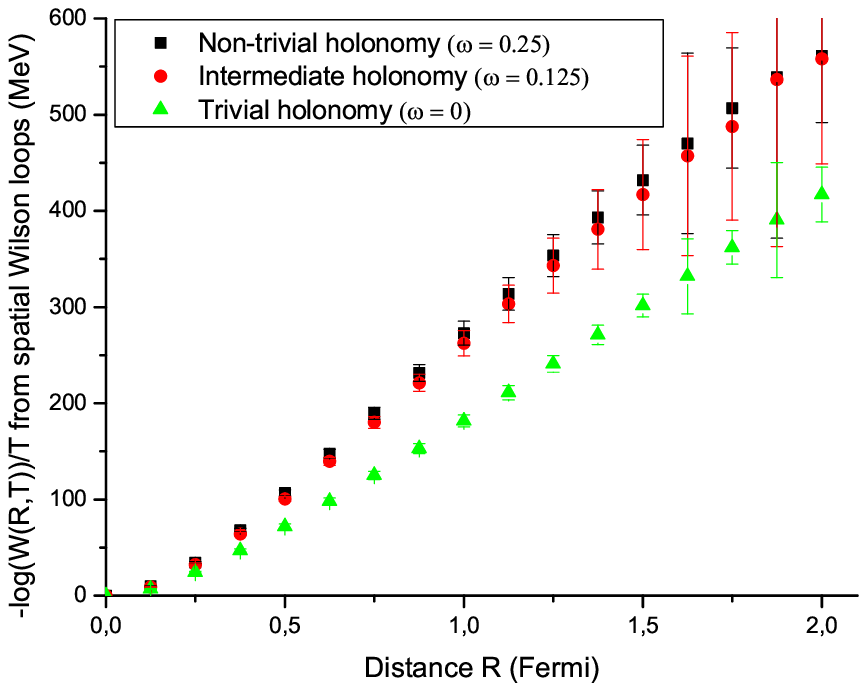}
\caption{Left: the potential for fundamental and adjoint static 
charges from the corresponding Polyakov loop correlators; 
right: linearly rising ``potential'' for  fundamental charges 
from spatial Wilson loops. Note the various asymptotic holonomies 
$\omega$.}
\label{fig:fig5}
\end{figure}
Though, the effect of non-trivial holonomy is striking.
Fig. \ref{fig:fig5} (left) shows that the static potential of fundamental
charges (shown as function of $R$) is strongly influenced by $\omega=1/4$, 
$1/8$ and $0$.  
Even for maximal non-trivial holonomy we observe string breaking for adjoint 
charges, whereas the initial rise of both potential corresponds to 
approximate Casimir scaling. 
Fig. \ref{fig:fig5} (right) presents $-\log W(R,T)/T$ from spatial Wilson 
loops demonstrating the existence of a spatial string tension even for trivial 
holonomy. 

So far, the temperature dependence of the holonomy 
$L_{\infty}$ and of the density $n_{\rm cal}$ was beyond the scope of 
our discussion. The complete model and its practical implementation 
(with realistic input distributions) will be described in the full
paper.
\vspace{-2mm}

\section{Summary}

We have updated our analysis to confirm KvB caloron-like 
correlations in smeared equilibrium configurations at various temperatures.
We reported on progress constructing a dyon-caloron gas model 
comprising holonomy, confinement and deconfinement in the vicinity of $T_{\rm dec}$. 
\vspace{-2mm}

\section*{Acknowledgments}
This work was supported by DFG (FOR 465 ``Lattice Hadron Phenomenology''. 
The ITEP--HU collaboration took benefit from grants RFBR 03-02-16941, 
RFBR 04-02-16079 and DFG 436 RUS 113/739/0. E.-M.~I. and B.~V.~M. thank 
for the hospitality in the group of P. van Baal at Leiden.
\vspace{-2mm}


\providecommand{\href}[2]{#2}\begingroup\raggedright\endgroup

\end{document}